\documentclass[final,twocolumn,prl,aps,floats,amsfonts,amssymb,showpacs,superscriptaddress]{revtex4}
\usepackage{amsmath}
\usepackage{graphicx}
\usepackage{color}
\usepackage{psfrag}

\makeatletter
\newlength{\earraycolsep}
\def\eqnarray{\stepcounter{equation}\let\@currentlabel%
\theequation
\global\@eqnswtrue\m@th
\global\@eqcnt\z@\tabskip\@centering\let\\\@eqncr
$$\halign to\displaywidth\bgroup\@eqnsel\hskip\@centering
$\displaystyle\tabskip\z@{##}$&\global\@eqcnt\@ne
\hskip 2\earraycolsep \hfil$\displaystyle{##}$\hfil
&\global\@eqcnt\tw@ \hskip 2\earraycolsep
$\displaystyle\tabskip\z@{##}$\hfil
\tabskip\@centering&\llap{##}\tabskip\z@\cr}
\makeatother
\begin{document}

\title{Fluctuation-Dissipation Relations and statistical temperatures in a turbulent von K\'arm\'an flow}

\author{Romain Monchaux}
\author{Pierre-Philippe Cortet}
\affiliation{Service de Physique de l'\'Etat Condens\'e, DSM, CEA
Saclay, CNRS URA 2464, 91191 Gif-sur-Yvette, France}
\author{Pierre-Henri Chavanis}
\affiliation{Laboratoire de Physique Th\'eorique, CNRS UMR 5152,
Universit\'e Paul Sabatier, 118 route de Narbonne,
31062 Toulouse, France}
\author{Arnaud Chiffaudel}
\author{Fran\c cois Daviaud}
\author{Pantxo Diribarne}
\author{B\'ereng\`ere Dubrulle}
\affiliation{Service de Physique de l'\'Etat Condens\'e, DSM, CEA
Saclay, CNRS URA 2464, 91191 Gif-sur-Yvette, France}


\begin{abstract}
We experimentally characterize the fluctuations of the
non-homogeneous non-isotropic turbulence in an axisymmetric von
K\'arm\'an flow. We show that these fluctuations satisfy
relations, issued from the Euler equation, which are analogous to
classical Fluctuation-Dissipation Relations in statistical
mechanics. We use these relations to estimate statistical
temperatures of turbulence.
\end{abstract}

\maketitle

Fluctuation-Dissipation Relations (FDRs) are one of the
corner-stone of statistical mechanics. They offer a direct
relation between the fluctuations of a system at equilibrium and
its response to a small external perturbation. Classical outcome
of FDRs are Einstein or Nyquist relations, or, more generally,
measures of the susceptibility, dissipation coefficient or
temperature of the system. The hypothesis behind the FDRs restrict
their applicability to systems that are close to equilibrium.
Theoretical extrapolation of the FDRs to systems far from
equilibrium is currently a very active area of research
\cite{cugliandolo93a}. In this context, experimental tests in
several glassy systems have evidenced violation of FDRs
\cite{buisson03a}. Furthermore, general identities about
fluctuations and dissipation, theoretically derived for
time-symmetric out of equilibrium systems \cite{evans93a}, have
been tested in dissipative (non time-symmetric) systems like
electrical circuit or turbulent flow
\cite{vanzon04aciliberto04}. Turbulence is actually a very special example
of far from equilibrium system. Due to its intrinsic dissipative
nature, an unforced turbulent flow is bound to decay to rest.
However, in the presence of a permanent forcing, a steady state
regime can be established, in which forcing and dissipation
equilibrate on average, allowing the maintenance of non-zero
averaged velocities, with large fluctuations covering a wide range
of scales. We use measurements performed in a
turbulent von K\'arm\'an flow to show that there is actually a
direct link between these fluctuations and the mean flow
properties, in a way analogous to classical FDRs. This approach
provides an estimate of effective statistical temperatures of our
turbulent flow.

\paragraph*{Theoretical background and definitions.-}
Describing turbulence with tools borrowed from statistical
mechanics is a long-standing dream, starting with Onsager
\cite{onsager49}. Advances in that direction have been recently
made for flows with symmetries (2D \cite{chavanis04}, axisymmetric
\cite{leprovost05}) using tools developed independently by Robert
and Sommeria and Miller \cite{sommeria91amiller90}. They
consider freely evolving flows described by the Euler equation (no
forcing and no dissipation). The Euler equation conserves the
energy and, for axisymmetric and shear flows, the helicity.  In
addition, owing to the symmetry, there is conservation of a local
scalar quantity along a velocity line (vorticity in 2D, angular
momentum for axisymmetry) resulting in a Liouville theorem and
additional global conserved quantities as Casimirs of the local
scalar quantity. In the Miller-Robert-Sommeria theory, the Euler
equation develops a mixing process leading to a quasi-stationary
state on the coarse-grained scale. This state is determined by the
initial conditions and maximizes a mixing entropy under
conservation of all the inviscid invariants. For forced
dissipative flows, the strict conservation of the inviscid
invariants is lost. However, an equilibrium between forcing and
dissipation can establish itself and the system can reach a steady
state that is a combination of a stationary solution of the Euler
equation and fluctuations. This steady state is selected by
forcing and dissipation. Some authors \cite{chavanis04,eht} have
proposed to describe this state by maximizing a mixing entropy
under only particular constraints (see for example $H_f$ and $I_g$
in Eq. \ref{gdeurglob2}) selected implicitly by forcing and
dissipation. This allows the derivation of Gibbs states of the
system from which one derives general identities characterizing
the steady states, as well as relations between these steady
states and their fluctuations.

We apply this approach to the axisymmetric case
which is relevant to our experimental device. In that case, the
variables describing the system are the angular momentum $\sigma$,
the stream function $\psi$ and the rescaled azimuthal vorticity
$\xi =r^{-1}\omega_\theta$ \cite{leprovost05,monchaux06a}. The
representation of the flow through $(\sigma, \psi, \xi)$ or
through the classical velocity components $(u_r,u_{\theta},u_z)$
in cylindrical coordinates ($r$,$\theta$,$z$) are equivalent since
$(u_r,0,u_z) = \nabla\times (r^{-1}\psi\,{\bf e}_\theta)$ and
$u_{\theta}=\sigma/r$. Furthermore,
$r^{-1}\partial_r\left(r^{-1}\partial_r
\psi\right)+r^{-2}\partial_z^2\psi=-\xi$. In the inviscid,
force-free limit, the global quantities conserved by the Euler
equation are the energy $E$, the generalized helicities $H_f$ and
the Casimirs $I_g$, given by \cite{leprovost05}:
\begin{eqnarray*}
{E} & = & \frac{1}{2} \int \xi \psi rdr dz + \frac{1}{2}\int
\frac{\sigma ^2}{r^2} rdr dz,
\end{eqnarray*}
\vskip -0.4cm
\begin{eqnarray}
{H_f} & = & \int \xi f(\sigma) rdr dz, \quad {I_g} = \int
g(\sigma) dy dz, \label{gdeurglob2}
\end{eqnarray}
where $f$ and $g$ are arbitrary functions. For forced dissipative
flows, it has been suggested to conserve only the energy  $E$ and
two particular integrals $H_{f}$ and $I_{g}$ where the functions
$f$ and $g$ are selected through forcing and dissipation
\cite{monchaux06a}. For Beltrami flows, in which vorticity is
proportional to velocity everywhere, i.e. ${\bf
u}=\lambda\,\nabla\times {\bf u}$, the relevant conserved
quantities are the energy $E$ and the helicity $H$ so that
$f(\sigma)=\sigma$ and $g=0$ \cite{leprovost05}.

Let us now apply the statistical mechanics approach introduced in
the previous paragraph to a Beltrami flow. The detailed procedure
is described in \cite{leprovost05,monchaux07c}. The mixing entropy
of the flow $S[\rho]$ is defined using the probability density
$\rho(\sigma,\xi,{\bf r})$ to have a certain couple of values for
$\sigma$ and $\xi$ at each position ${\bf r}$. Because of the
Beltrami hypothesis, the steady state of the flow maximizes
$S[\rho]$ at fixed energy $E$ and helicity $H$. Then, writing
$\delta S-\beta\delta E-\mu\delta H=0$ (where $\beta^{-1}$ is a
temperature and $\mu$ an ``helical potential'') and using two
different mean field approximations, one finds two relations for
the averaged fields\footnote{$\overline{x}$ stands for statistical
average of $x$. Its experimental estimate is made by
time-averaging assuming ergodicity.}:
\vskip -0.4cm
\begin{subequations}
\label{eqstat}
\begin{eqnarray}
\beta_\xi\overline{\psi} + \mu_\xi\overline{\sigma}=0,
\label{eqstata}\\
\frac{\beta_\sigma \overline{\sigma}} {r^2} +
\mu_\sigma \overline{\xi}=0, \label{eqstatb}
\end{eqnarray}
\end{subequations}
\vskip -0.2cm
as well as two expressions for the probability density
$\rho_{\xi}(\sigma,{\bf r})$ and $\rho_{\sigma}(\xi,{\bf r})$,
which happen to be Gaussian functions. The two mean field
approximations consist in fixing independently $\xi$ or $\sigma$
to their time average. The thermodynamic coefficients $\mu$ and
$\beta$ have been labelled accordingly. Considering the first
moment of $\rho_{\xi}$, one gets an additional relation for the
averaged fields:
\vskip -0.4cm
\begin{equation}
\frac{\beta_\xi \overline{\sigma}}{r^2}+\mu_\xi\overline{\xi}=0.
\label{eqstatc}
\end{equation}
\vskip -0.2cm
Finally, considering the second moment of $\rho_{\xi}(\sigma,{\bf
r})$ and $\rho_{\sigma}(\xi,{\bf r})$, we obtain relations for
fluctuations:
\vskip -0.4cm
\begin{eqnarray}\label{FDR}
\overline{\sigma^2}-\overline{\sigma}^2 = \frac{r^2}{\beta_{\xi}},
\qquad  \overline{\xi ^2}-\overline{\xi
}^2=\frac{\beta_{\sigma}}{\mu_{\sigma}^2 r^2}.
\end{eqnarray}
\vskip -0.2cm
Comparing Eq. (\ref{eqstatb}) with Eq. (\ref{eqstatc}), we see
that the four thermodynamic coefficients in these equations obey:
\vskip -0.4cm
\begin{equation}\label{stationnarity}
\frac{\mu_\sigma}{\beta_{\sigma}}=\frac{\mu_\xi}{\beta_{\xi}}.
\end{equation}
\vskip -0.2cm
Dividing Eq. (\ref{eqstata}) by $r$ and taking its curl, we obtain
a relation between the poloidal components, i.e. the components in
$r$-$z$ plane, of the velocity and vorticity: ${\bf
\overline{u_p}}=-\mu_{\xi}/\beta_{\xi}\,${\boldmath$
\overline{\omega_{\rm p}}$}. Then, multiplying Eq. (\ref{eqstatb})
by $r$, we obtain the following relation for the toroidal, i.e.
azimuthal, components: {\boldmath $
\overline{u_\theta}$}$=-\mu_{\sigma}/\beta_{\sigma}\, ${\boldmath$
\overline{\omega_{\theta}}$}. Finally, using Eq.
(\ref{stationnarity}), we get:
\vskip -0.4cm
\begin{equation}\label{betrami}
{\bf
\overline{u}}=-\frac{\mu_{\sigma}}{\beta_{\sigma}}\,\nabla\times{\bf
\overline{u}},
\end{equation}
\vskip -0.2cm
and verify that the averaged flow is a Beltrami flow with a
constant
$\lambda=-\mu_\sigma/\beta_{\sigma}=-\mu_\xi/\beta_{\xi}$. Note
that this flow minimizes the coarse-grained energy at fixed
helicity. Additionally, combining Eqs. (\ref{eqstata}) and
(\ref{eqstatb}), we find that:
\vskip -0.4cm
\begin{equation}
{\bf\overline{u_p}}=\frac{\mu_\sigma}{\beta_{\sigma}}\frac{\mu_\xi}{\beta_{\xi}}\,
\nabla \times (\overline{\omega _{\theta}}\,{\bf e}_\theta
)=\lambda ^{2}\,\nabla \times (\overline{\omega _{\theta}}\,{\bf
e}_\theta ), \label{fickvorticity}
\end{equation}
\vskip -0.2cm
namely that a spatial variation of the azimuthal vorticity creates
a poloidal velocity with a proportionality constant $\lambda^2$.
This relation can be used to define a susceptibility $\chi\equiv
1/\lambda^2$ which happens to be always positive.

Eqs. (\ref{FDR}) can easily be recast into:
\vskip -0.4cm
\begin{subequations}
\label{FDRbis}
\begin{eqnarray}
\label{FDRbisun}
\overline{u_\theta^2}-\overline{u_\theta}^2 &=& \frac{1}{\beta_{\xi}},\\
\label{FDRbisdeux}
\overline{\omega_\theta^2}-\overline{\omega_\theta
}^2&=&\frac{\chi}{\beta_{\sigma}},
\end{eqnarray}
\end{subequations}
\vskip -0.2cm
predicting uniformity of azimuthal velocity and vorticity
fluctuations. Eq. (\ref{FDRbisun}) shows that the azimuthal
velocity fluctuations define an effective statistical temperature
$1/\beta_{\xi}$. Eq. (\ref{FDRbisdeux}) links the vorticity
fluctuations to the susceptibility $\chi$ and a vortical effective
temperature $1/\beta_{\sigma}$. This is a formal equivalent of the
Einstein relation for the Brownian motion. These equations may be
regarded as formally analogous to FDRs since they link
fluctuations, susceptibility and temperature. These predictions
enable the measurements of turbulence effective temperatures
through fluctuations of $u_{\theta}$ and $\omega_{\theta}$ in a
Beltrami flow. Because variances are positive, $\beta_\sigma$ and
$\beta_\xi$ are always positive (since $\chi>0$), unlike in the 2D
situation where $\chi$ can be negative 
\footnote{For a statistical
equilibrium state of the 2D Euler equation, with general
$\overline{\omega}=f({\psi})$ relationship, one can
derive $\overline{\omega^{2}}-\overline{\omega}^{2}=(1/\beta)
d\overline{\omega}/d{\psi}$ \cite{chavanis04} which is analogous
to a FDR since it links the fluctuations of $\omega$ to the
temperature $1/\beta$ and the susceptibility $\chi=d
\overline{\omega}/d{\psi}$. This relation is the counterpart of
Eq. (\ref{FDRbisdeux}). 2D velocity ${\bf
v}=\nabla\times (\psi {\bf e}_{z})$ can be written ${\bf
v}=\chi^{-1}\nabla\times (\overline{\omega}{\bf e}_{z})$ which is
the counterpart of Eq. (\ref{fickvorticity}).}. 
In contrast,
$\mu_\sigma$ and $\mu_\xi$ can take positive or negative values,
depending on the helicity sign.

The analogy between our predictions and FDRs can actually be
pushed forward. An other possible way to derive Eqs.
(\ref{FDR}) is to introduce, as in classical statistical
mechanics, the partition function $Z_{\xi,\sigma}$ describing the
Beltrami equilibrium state in each mean field approximation:
\vskip -0.4cm
\begin{subequations}
\label{FDRter}
\begin{eqnarray}
\overline{\sigma^2}-\overline{\sigma}^2 &=&
\frac{1}{\mu_{\xi}^2}\frac{\delta^2 \log Z_\xi}{\delta
\overline{\xi}^2}= -\frac{1}{\mu_{\xi}}\frac{\delta
\overline{\sigma}}{\delta \overline{\xi}},\label{FDRterun}\\
\overline{\xi ^2}-\overline{\xi }^2&=& \frac{1}{\mu_{\sigma
}^2}\frac{\delta^2 \log Z_\sigma}{\delta \overline{\sigma}^2}=
-\frac{1}{\mu_{\sigma}}\frac{\delta \overline{\xi}}{\delta
\overline{\sigma}}.\label{FDRterdeux}
\end{eqnarray}
\end{subequations}
\vskip -0.2cm
Formally, the mathematical objects $\delta
\overline{\sigma}/\delta \overline{\xi}$ and $\delta
\overline{\xi}/\delta \overline{\sigma}$ can be seen as response
functions. With this point of view, Eqs. (\ref{FDRter}) again
reflect a formal analogy with FDRs since another classical way to
write it down is to link the fluctuations of a field to its
response to a perturbation.

The challenge is to face such fluctuation
relations with experimental data. For this, we use a turbulent von
K\'arm\'an flow that has already been shown to tend to a Beltrami
flow at large Reynolds number \cite{monchaux06a}.

\paragraph*{Experimental flow and measurement techniques.-}
Our experimental setup consists of a plexiglas cylinder (radius
$R_c=100$ mm) filled up with water. The fluid is mechanically
stirred by a pair of impellers driven by two independent motors in
exact counterrotation. The resulting flow belongs to the von
K\'arm\'an class of flows with a mean flow divided into two toric
cells separated by an azimuthal shear layer. We define the
Reynolds number as $Re=UL/\nu= 2\pi F R_c^2/\nu$, where $F$ is the
impeller frequency and $\nu$ the water viscosity. Rotating the
impellers from $2$ to $8$Hz, we can achieve Reynolds numbers from
$125,000$ to $500,000$. Our two models of impellers, TM60 and
TM73, are flat disks of respective radius $0.925\,R_c$ and
$0.75\,R_c$, fitted with radial blades of height $0.2\,R_c$ and
respective curvature $0.50\,R_c$ and $0.925\,R_c$. The inner face
of the discs are $1.8\,R_c$ appart. Different forcings are
associated with the convex or concave face of the blades going
forward, denoted in the sequel by senses ($+$) and ($-$). Velocity
measurements are done with a Stereoscopic Particule Image
Velocimetry system \cite{monchaux07c} provided by DANTEC Dynamics.
The cylinder is mounted inside a water filled square plexiglas
container in order to reduce optical deformations. Two digital
cameras are aiming at a meridian plane of the flow through two
perpendicular faces of the square container giving a 2D-three
components velocity field map. Correlation calculations are done
on $32\times 32$ pixels$^2$ windows with $50\%$ overlap. As a
result, each velocity is averaged on a $4.16 \times 4.16$~mm$^2$
window over the $1.5$~mm laser sheet thickness. The spatial
resolution is $2.08$~mm. A basic measurement is a set of $5000$
acquisitions at a rate of $4$ images per second. From this set, we
compute time-average and fluctuations of the three velocity
components.

From now on, all physical quantities and equations will be
considered in their non-dimensionalized form using $F$ and $R_c$.
For each of the four forcing configurations, we made between five
and seven tests that show no $Re$-dependency for these
non-dimensionalized quantities.

\paragraph*{Test of the mean flow relations.-}

\begin{figure}
\psfrag{p}[l][][1.2]{$\overline{\psi}$\tiny{$\times
10^{-2}$}}\psfrag{s}[c][][1.2]{$\overline{\sigma}$}
\psfrag{x}[l][][1.2]{$\overline{\xi} r^2$}
\centerline{\includegraphics[width=8.2cm]{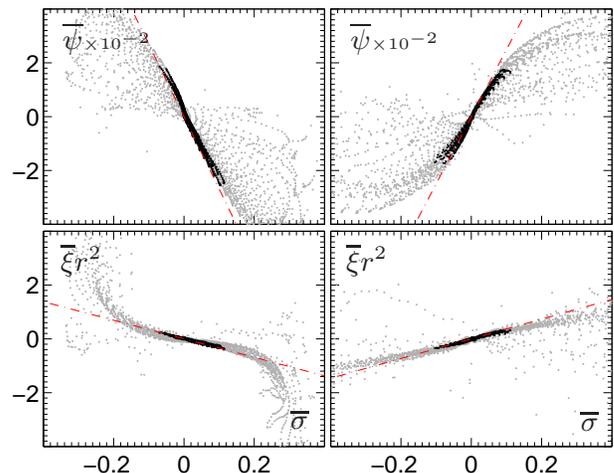}}
\caption{$\overline\psi$ (top) and $\overline{\xi} r^2$ (bottom)
versus $\overline\sigma$ for two experimental von K\'arm\'an flow
with TM60 impellers at $F=6$Hz, sense $(+)$ at left, $(-)$ at
right. Black dots correspond to flow bulk data ($|z| \leq 0.5$,
$|r|\leq 0.5$) and define mostly linear functions. The slopes of
the dot-dashed lines is given by the first order coefficient of an
odd cubic fit of the data. Corresponding values of
$\beta_{\sigma}/\mu _{\sigma}$ and $\beta _{\xi}/\mu _{\xi}$ are
used to compute Table \ref{sumupres}.} \label{stat}
\end{figure}

Relations (\ref{eqstata}) and (\ref{eqstatb}) can be tested by
plotting $\overline\psi$ and $\overline{\xi} r^2$ with respect to
$\overline{\sigma}$ for each experiment. Two of them are plotted
in Fig.~\ref{stat}. As in Ref. \cite {monchaux06a}, we focus on
the flow bulk, i.e. $|z| \leq 0.5$ and $|r|\leq 0.5$, where the
von K\'arm\'an flow is close to a Beltrami flow. Linear
dependencies of Eqs. (\ref{eqstat}) are confirmed \cite
{monchaux06a} and enable estimation of the slopes
$\beta_{\sigma}/\mu _{\sigma}$ and $\beta _{\xi}/\mu _{\xi}$ (see
caption of Fig.~\ref{stat} for details and Table \ref{sumupres}
for averaged results). Depending on the forcing, the two
measurements of $\beta/\mu$ differ from $1$ to $13\%$, verifying
Eq.~(\ref{stationnarity}) and giving a unique mean value of
$\beta/\mu$ or $\lambda$ (see Table \ref{sumupres}, line 3). The
quality of the test of Eq.~(\ref{stationnarity}) is evaluated by
$\beta_{\xi}/\mu _{\xi}-\beta_{\sigma}/\mu_{\sigma}$ (line 4). One
can see that the two $68\%$ confidence intervals of
$\beta_{\xi}/\mu _{\xi}$ and $\beta_{\sigma}/\mu_{\sigma}$ always
overlap.

\begin{figure}
\psfrag{u}[l][][1.2]{$\overline{u_\theta^2}-\overline{u_\theta
}^2$}\psfrag{w}[l][][1.2]{$\overline{\omega_\theta^2}-\overline{\omega_\theta
}^2$} \psfrag{r}[c][][1]{$r$}\centerline{
\includegraphics[width=8.2cm]{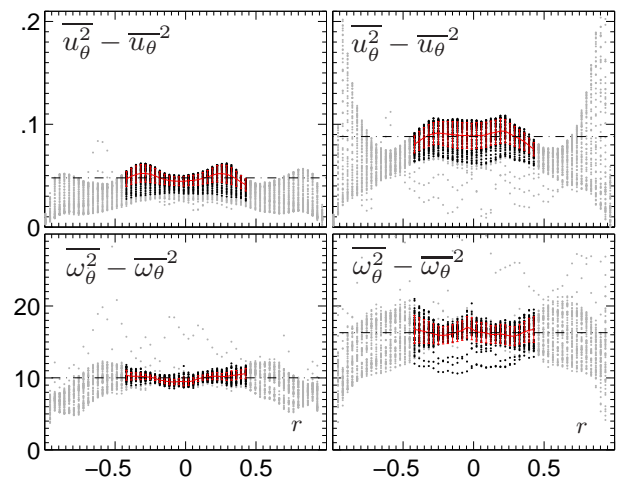}}
\caption{Evolution of angular velocity fluctuations (top) and
angular momentum fluctuations (bottom) with the radial coordinate
$r$ for the same flows as in Fig. \ref{stat}. Black dots
correspond to flow bulk data ($|z| \leq 0.5$, $|r|\leq 0.5$). The
corresponding mean values and standard deviations at each $z$ are
plotted by red lines and errorbars. Horizontal dot-dashed lines
show the averages over the flow bulk, i.e. measured values for
$1/\beta_{\xi}$ and $\beta_{\sigma} / \mu_{\sigma}^2$. }
\label{fluct}
\end{figure}
\begin{table*}[ht]
\begin{tabular}{cc}
\parbox{12cm}{
\begin{tabular}
{||c||c|c||c|c||}%
\hline \hline
    Impellers   & \multicolumn{2}{|c||}{TM73}& \multicolumn{2}{|c||}{TM60}\\
\hline \hline
    Sense   &$  (+)         $&$ (-)         $&$ (+)         $&$ (-)         $\\
\hline \hline
$   \beta_{\xi} / \mu _{\xi}    $&$ 4.64    \pm 0.25    $&$ -4.92   \pm 0.12    $&$ 3.76    \pm 0.28    $&$ -4.11   \pm 0.31    $\\
\hline
$   \beta_{\sigma} /\mu_{\sigma}    $&$ 4.31    \pm 0.20    $&$ -4.88   \pm 0.17    $&$ 3.55    \pm 0.20    $&$ -3.61   \pm 0.23    $\\
\hline
$   < \beta /\mu >  $&$ 4.47    \pm 0.22    $&$ -4.90   \pm 0.15    $&$ 3.66    \pm 0.24    $&$ -3.86   \pm 0.27    $\\
\hline
$   \beta_{\xi}/\mu _{\xi}-\beta_{\sigma}/\mu_{\sigma}  $&$ 0.33    \pm 0.45    $&$ -0.04   \pm 0.29    $&$ 0.21    \pm 0.48    $&$ -0.50   \pm 0.54    $\\
\hline
$   1/ \beta _{\xi} $&$ 0.0452  \pm 0.0040  $&$ 0.0673  \pm 0.0035  $&$ 0.0481  \pm 0.0056  $&$ 0.0922  \pm 0.0086  $\\
\hline
$   \beta_{\sigma} / \mu_{\sigma}^2 $&$ 9.1 \pm 0.5 $&$ 13.4    \pm 0.7 $&$ 10.4    \pm 1.4 $&$ 16.4    \pm 0.9 $\\
\hline
$   \beta_{\xi} $&$ 22.1    \pm 2.0 $&$ 14.9    \pm 0.8 $&$ 20.8    \pm 2.4 $&$ 10.8    \pm 1.0 $\\
\hline
$   \beta_{\sigma}  $&$ 2.04    \pm 0.31    $&$ 1.77    \pm 0.21    $&$ 1.22    \pm 0.31    $&$ 0.79    \pm 0.14    $\\
\hline
$   \mu_{\xi}   $&$ 4.77    \pm 0.68    $&$ -3.02   \pm 0.23    $&$ 5.53    \pm 1.06    $&$ -2.64   \pm 0.45    $\\
\hline
$   \mu_{\sigma}    $&$ 0.47    \pm 0.05    $&$ -0.36   \pm 0.03    $&$ 0.34    \pm 0.07    $&$ -0.22   \pm 0.03    $\\
\hline \hline
\end{tabular}
}
&\parbox{6cm}{
{\caption{Non-dimensionalized statistical coefficients measured in
our experiments following Eqs. (\ref{eqstat}) and (\ref{FDRbis})
for the four forcings studied (two impellers, two senses). Raw
measurements (lines 1-2 and 5-6) allow to calculate a Beltrami
factor (line 3-4) and statistical coefficients (lines 7-10).
Errors are calculated with standard $68\%$ confidence intervals as
the sum of the error on data fits (see Figs. \ref{stat} and
\ref{fluct}) and of the statistical dispersion over the different
runs. Each measurement is an average over $5$ to $7$ experiments
performed at high $Re$ between $1.2$ and $5\times 10^5$.}
\label{sumupres}}
}
\end{tabular}
\end{table*}

\paragraph*{Test of the fluctuation relations.-}
Now, we turn to experimental test of fluctuation relations
(\ref{FDRbis}) and to complete determination of the four \emph{a
priori} independent coefficients: $\beta_\xi$, $\beta_\sigma$,
$\mu_\xi$ and $\mu_\sigma$. Because of Eq.~(\ref{stationnarity}),
only three of them are independent and the previous test already
provided a measurement of $\beta/\mu$. So, we can use the
fluctuation relations of Eqs. (\ref{FDRbis}) to compute the
remaining parameters $1/\beta_{\xi}$ and
$\beta_{\sigma}/\mu_{\sigma}^2$.

Fig. \ref{fluct} presents the analysis of the fluctuation
relations for angular velocity $u_{\theta}$ and vorticity
$\omega_{\theta}$. On the top plots, we test relation
(\ref{FDRbisun}). Over the whole flow, the velocity fluctuations
are roughly constant. The relative scattering, which mostly tracks
the $z$-dependance, increases with $r$. Focusing again on the flow
bulk, we observe a reduced scattering and measure $1/\beta _{\xi}$
with a simple average. On the bottom plots, we perform the same
analysis for vorticity fluctuations and relation
(\ref{FDRbisdeux}). The vorticity fluctuations displays a
scattering of the same order of magnitude than angular velocity
fluctuations. Once again, an average over the bulk allows to
measure $\beta_{\sigma}/\mu _{\sigma}^2$. We conclude that, for
these von K\'arm\'an flow, the variances of azimuthal velocity and
vorticity are constant in the flow bulk. This result is compatible
with the mean field analysis drawn in the first part of this
Letter for a Beltrami flow. Combining the two sets of results
described above, we can independently evaluate ($\mu _{\sigma}$,
$\beta _{\sigma}$) and ($\mu _{\xi}$, $\beta _{\xi}$). Standard
error estimates are typically $10\%$, ranging from $5$ to $20\%$.
Table \ref{sumupres} shows that these two couples are not equal to
each other and even differ by one order of magnitude. This result
is discussed in the next section.

\paragraph*{Conclusion.-}

We have derived predictions concerning the mean
flow and the fluctuations in an axisymmetric Euler-Beltrami flow
using tools borrowed from statistical mechanics. These
predictions, and especially the uniformity of the variances of
$u_\theta$ and $\omega_\theta$ (Eqs. \ref{FDRbis}), have then been
experimentally confirmed in the specific case of a turbulent von
K\'arm\'an flow at large Reynolds number. This result is a priori
unexpected since the theory, issued from the Euler equation, does
not explicitly take into account the forcing and the dissipation
which {\it implicitly} determine the form of the steady state.
This is an additional confirmation that out of equilibrium steady
states of a real turbulent flow may be described as equilibrium
states of the Euler equation as suggested in \cite{monchaux06a}.
Additionally, these relations provide two different values for an
effective statistical temperature $1/\beta$ of our system
depending on the selected variable. Such non-uniqueness of
statistical temperature in out of equilibrium systems has already
been encountered in the context of glassy systems
\cite{Fielding02}. Turbulent flows may be another example of this
out of equilibrium property. Meanwhile, one could possibly use
correlation with hydrodynamics properties (such as variation with
forcing, fluctuation rate or Reynolds number \cite{ravelet08}) 
to decide whether one of the temperatures we measure is more relevant
than the other to stand for a statistical temperature of the
turbulence. Finally, in order to test further the physical relevance of the
FDR analogy, one should perform direct measurements of response
functions as suggested by the \emph{r.h.s.} of Eqs.
(\ref{FDRter}). This is a yet unresolved and challenging
experimental issue.

This work was supported by ANR TSF NT05-1-41492.

\end{document}